\documentclass[nofootinbib,twocolumn,showpacs,preprintnumbers,superscriptaddress,prb,floatfix,aps,10pt]{revtex4-1}

\usepackage{graphicx}
\usepackage{textcomp}  
\usepackage{natbib}
\usepackage{xcolor}

\begin{document}

\title{Local Photothermal Control of Phase Transitions for \\ On-demand Room-temperature Rewritable Magnetic Patterning}
\author{Antonio B. Mei} \email{amei2@illinois.edu} \affiliation{Department of Materials Science and Engineering, \\ Cornell University, Ithaca, NY, 14853, USA}
\author{Isaiah Gray} \affiliation{School of Applied and Engineering Physics, Cornell University, Ithaca, NY, 14853, USA}
\author{Yongjian Tang} \affiliation{Department of Physics, Cornell University, Ithaca, NY, 14853, USA}
\author{Jü\"urgen Schubert} \affiliation{Peter Gr\"unberg Institute (PGI-9) and JARA-Fundamentals of Future Information Technology, \\ Forschungszentrum J\"ulich GmbH, 52425 J\"ulich, Germany}
\author{Don Werder} \affiliation{Department of Materials Science and Engineering, \\ Cornell University, Ithaca, NY, 14853, USA} \affiliation{Cornell Center for Material Research, \\ Cornell University, Ithaca, NY, 14853, USA}
\author{Jason Bartell} \affiliation{School of Applied and Engineering Physics, Cornell University, Ithaca, NY, 14853, USA}
\author{Daniel C. Ralph} \affiliation{Department of Physics, Cornell University, Ithaca, NY, 14853, USA} \affiliation{Kavli Institute at Cornell for Nanoscale Science, Ithaca, NY, 14853, USA}
\author{Gregory D. Fuchs} \affiliation{School of Applied and Engineering Physics, Cornell University, Ithaca, NY, 14853, USA}
\author{Darrell G. Schlom} \affiliation{Department of Materials Science and Engineering, \\ Cornell University, Ithaca, NY, 14853, USA} \affiliation{Kavli Institute at Cornell for Nanoscale Science, Ithaca, NY, 14853, USA}

\maketitle

\textbf{The ability to make controlled patterns of magnetic structures within a nonmagnetic background is essential for several types of existing and proposed technologies. Such patterns provide the foundation of magnetic memory and logic devices\cite{Imre:2006is}, allow the creation of artificial spin-ice lattices\cite{Wang:2006hea,Louis:2018bh} and enable the study of magnon propagation\cite{Chumak:2015fa}. Here, we report a novel approach for magnetic patterning that allows repeated creation and erasure of arbitrary shapes of thin-film ferromagnetic structures. This strategy is enabled by epitaxial Fe\textsubscript{0.52}Rh\textsubscript{0.48} thin films designed so that both ferromagnetic and antiferromagnetic phases are bistable at room temperature. Starting with the film in a uniform antiferromagnetic state, we demonstrate the ability to write arbitrary patterns of the ferromagnetic phase by local heating with a focused laser. If desired, the results can then be erased by cooling with a thermoelectric cooler and the material repeatedly re-patterned.}

Intermetallic Fe\textsubscript{1-$x$}Rh\textsubscript{$x$} (B2, $Pm\bar{3}m$) exhibits a hysteretic antiferromagnetic/ferromagnetic transformation which has been harnessed to produce composite multiferroics exhibiting record-breaking magnetoelectric coupling coefficients\cite{Cherifi:2014du}, magnetocaloric refrigerators with competitive cooling capabilities\cite{Liu:2016kl} and novel memories based on antiferromagnetic order\cite{Marti:2014fl}. In this study, we design epitaxial Fe\textsubscript{1-$x$}Rh\textsubscript{$x$} films so that both ferromagnetic and antiferromagnetic states are simultaneously bistable at room temperature. We engineer the width of the thermal hysteresis to be sufficiently narrow to enable efficient controllability, but also wide enough to robustly withstand thermal perturbations. Moderate heating by a focused laser is used to locally drive antiferromagnetic regions controllably to the ferromagnetic phase, demonstrating the patterning of arbitrary magnetic features on the submicron scale.  These findings present opportunities for writing and erasing high-fidelity magnetically active nanostructures that are of interest for magnonic crystals\cite{Vogel:2015ky}, artificial spin-ice lattices\cite{Nisoli:2013dd} and memory\cite{Weller:2014fh} and logic devices\cite{Cowburn:2000to}. 

\begin{figure}[!bt]
\includegraphics[width=0.47\textwidth]{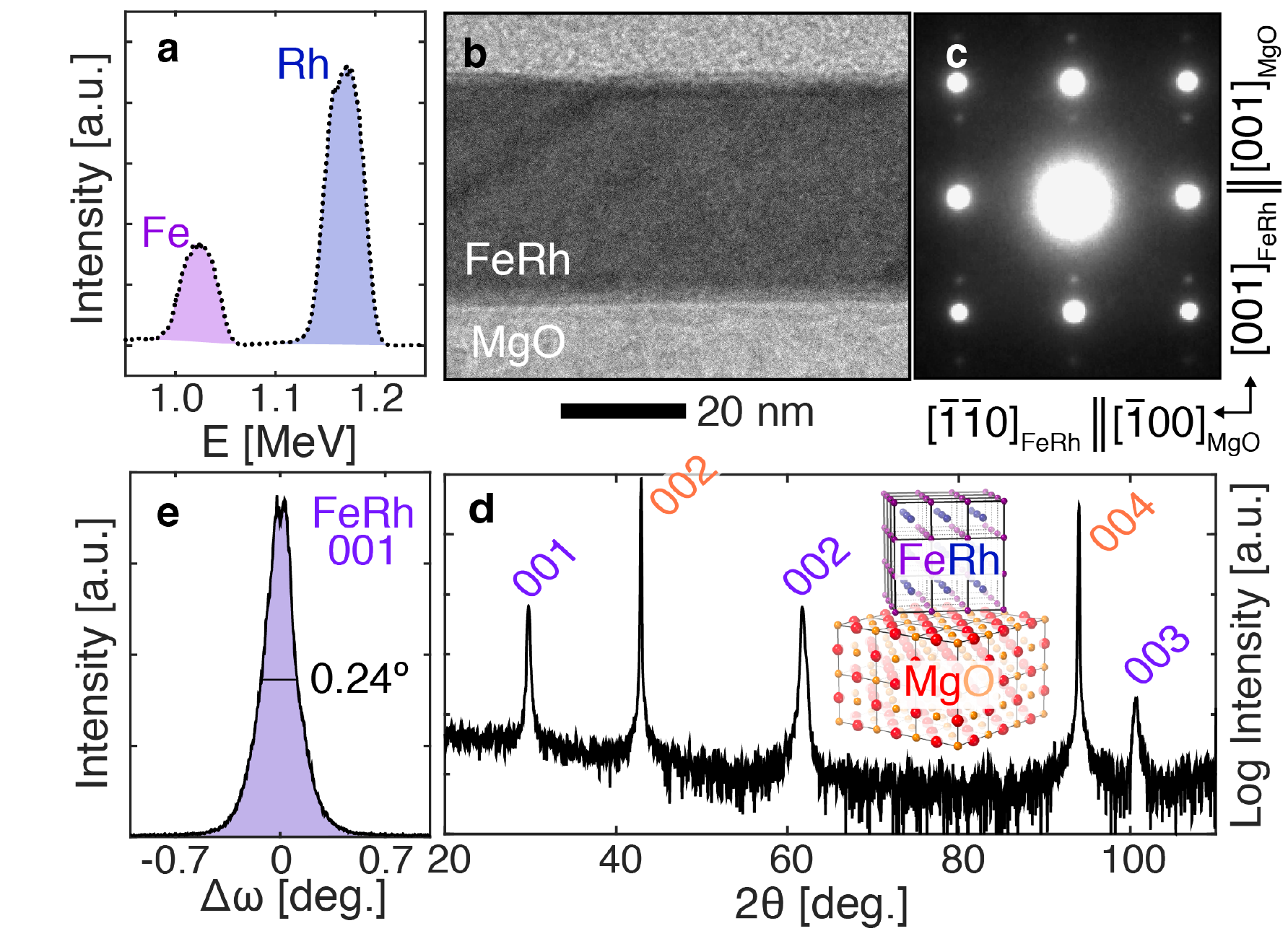}
\caption{\label{fig:sample}
\textbf{Fully-dense phase-pure untwinned epitaxial B2 Fe\textsubscript{0.52}Rh\textsubscript{0.48}/MgO(001) layers grown via molecular-beam epitaxy.}
\textbf{a}, RBS spectrum; labeled iron and rhodium spectral features correspond to a magnetically bistable rhodium fraction of 0.48.
\textbf{b}, BF-TEM image of the entire film cross section together with \textbf{c}, a corresponding SAED pattern. Note the weaker film and more intense substrate reflections.
\textbf{d}, XRD $\theta$-$2\theta$ scan; film (violet) and substrate (orange) reflections are indexed.
\textbf{e}, XRD $\omega$-rocking scan about the 001 film peak.
 }
\end{figure}

Fe\textsubscript{1-$x$}Rh\textsubscript{$x$} films are grown epitaxially on single-crystalline (001)-oriented MgO substrates using molecular-beam epitaxy. The fraction $x$ of rhodium in the film is carefully tuned to 0.48, where the hysteretic antiferromagnet/ferromagnet phase transitions are centered near room temperature. Film compositions are confirmed via Rutherford backscattering spectrometry (RBS) measurements (Fig. \ref{fig:sample}\textbf{a}) using the areal ratio of corresponding iron and rhodium spectral features. Cross-sectional bright-field transmission electron microscopy (BF-TEM) images (Fig. \ref{fig:sample}\textbf{b}), selected area electron diffraction (SAED) patterns (Fig. \ref{fig:sample}\textbf{c}) and x-ray diffraction (XRD) $\theta$-$2\theta$ scans (Fig. \ref{fig:sample}\textbf{d}) demonstrate that the films are fully-dense phase-pure untwinned epitaxial Fe\textsubscript{0.52}Rh\textsubscript{0.48} layers with the B2 CsCl-structure and that the film lattice is rotated 45$^\circ$ in-plane with respect to the underlying B1 NaCl-structure MgO(001) substrate crystal: (001)\textsubscript{Fe\textsubscript{0.52}Rh\textsubscript{0.48}} $\mid\mid$ (001)\textsubscript{MgO} and [110]\textsubscript{Fe\textsubscript{0.52}Rh\textsubscript{0.48}} $\mid\mid$ [100]\textsubscript{MgO}. The structural perfection of the films is established from the width of the $\omega$-rocking curve scans (Fig. \ref{fig:sample}\textbf{e}) to be consistent with our previous report of high-quality epitaxial Fe\textsubscript{1-$x$}Rh\textsubscript{$x$} layers\cite{Mei:2018dj}.

\begin{figure}[!bt]
\includegraphics[width=0.47\textwidth]{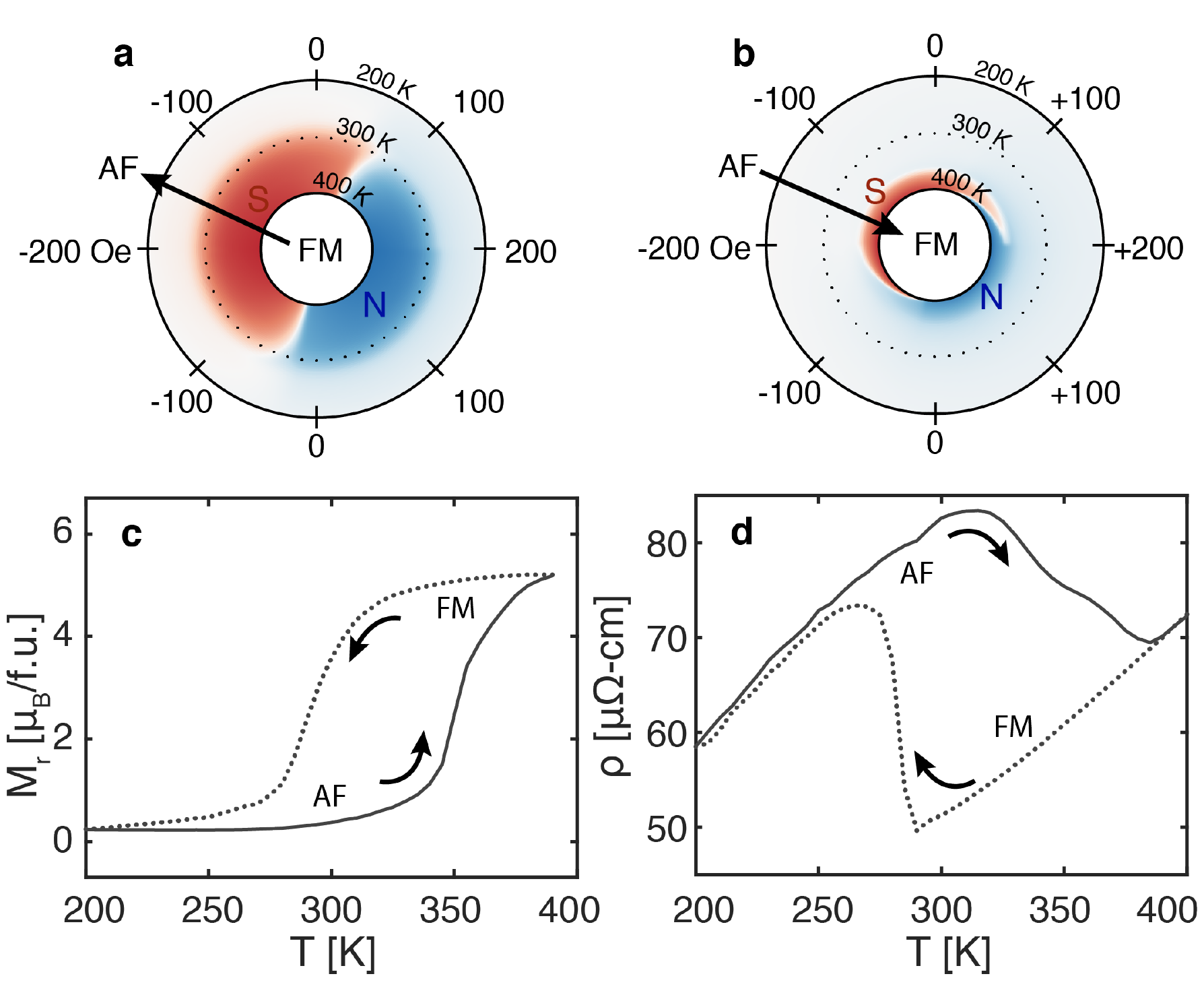}
\caption{\label{fig:MH}
\textbf{Designed room-temperature magnetic bistability of epitaxial Fe\textsubscript{0.52}Rh\textsubscript{0.48}/MgO(001) layers.}
In-plane magnetisation $\vec{M}$ as a function of $\vec{H}$ field (clockwise) and temperature $T$ (radial) during \textbf{a} cooling and \textbf{b} heating. Positive field points along Fe\textsubscript{0.52}Rh\textsubscript{0.48} $[110]$. Red and blue regions correspond to opposite orientations of the film magnetisation.
\textbf{c}, Temperature-dependent remanent magnetisation $M_r(T)$ and \textbf{d} resistivity $\rho(T)$ during cooling (dotted) and heating (solid).
}
\end{figure}

The room-temperature magnetic bistability of our epitaxial Fe\textsubscript{0.52}Rh\textsubscript{0.48}/MgO(001) layers is established through the combination of in-plane vibrating sample magnetometry (VSM) and electrical transport measurements. Magnetisation measurements collected as a function of temperature $T$ and applied magnetic field $\vec{H}$ during cooling and heating are presented in Fig. \ref{fig:MH}\textbf{a},\textbf{b}; $\vec{H}$ is applied along Fe\textsubscript{0.52}Rh\textsubscript{0.48} $[110]$. At room temperature, the as-deposited films exhibit a field-dependent magnetic hysteresis characteristic of FM order (Fig. \ref{fig:MH}\textbf{a}) with measured saturation magnetisations $M_s \approx  4 \mu_B/$f.u. and coercive fields $H_c \approx 50$ Oe, consistent with prior reports\cite{Mei:2018dj}.

 Cooling from 300 to 275 K suppresses the hysteresis associated with the FM state (Fig. \ref{fig:MH}\textbf{a}). In parallel, we observe a four-fold reduction in remanent magnetisation (Fig. \ref{fig:MH}\textbf{c}) from $M_r = $ 4 to 1 $\mu_B/$f.u. and an approximate 50\% increase in film resistivity (Fig. \ref{fig:MH}\textbf{d}) from $\rho = $ 50 to 72 $\mu\Omega$-cm. These features are consistent with a magnetic transition, in which initially ferromagnetic Fe\textsubscript{0.52}Rh\textsubscript{0.48} layers adopt an antiferromagnetic configuration characterised by anti-aligned neighbouring iron moments\cite{Shirane:1963bp}, decreased carrier densities\cite{Mankovsky:2017dta} and increased scattering rates\cite{Kudrnovsky:2015bz}. With decreasing $T$ below 250 K, remanent magnetisations decrease and saturate at $\approx 0.2 \mu_B/$f.u. (Fig. \ref{fig:MH}\textbf{c});  nonvanishing $M_r$ values are a common feature in Fe\textsubscript{1-$x$}Rh\textsubscript{$x$} films and have been attributed to remanent ferromagnetism near the film/substrate interface\cite{Fan:2010gu}.

 On heating to 350 K, the field-dependent magnetic hysteresis (Fig. \ref{fig:MH}\textbf{b}) and large remanent magnetisation values (Fig. \ref{fig:MH}\textbf{c}) defining the FM state are restored. The dissimilar critical temperatures during heating ($\approx$ 350 K) and cooling ($\approx$ 275 K) are a hallmark of a first-order phase transition and result in a thermal hysteresis and a window of bistability over which both FM and AF states can simultaneously coexist. As a result, not only can the magnetic order of our epitaxial Fe\textsubscript{0.52}Rh\textsubscript{0.48} layer be toggled between the AF and FM states using a commercial single-stage thermoelectric device, but the magnetic phase so defined will persist robustly at room temperature.
 
 As the temperature is further raised from 350 to 400 K, the coercive field of the re-established FM state decreases from $H_c \sim$ 200 to 50 Oe.  We attribute the enhanced coercivity at 350 K to exchange-coupling between recently-formed FM regions and the AF bulk, which decreases as the FM domains coalesce and the AF regions shrink at higher temperature. Over the same temperature range, resistivity values gradually decrease as a larger fraction of the sample transitions to the more conductive state associated with FM order (Fig. \ref{fig:MH}\textbf{d}). The gradual descent in film resistivity upon heating contrasts sharply with the abrupt jump observed upon cooling and reflects different kinetics in the heating and cooling branches of the transition, analogous to the asymmetry between melting and freezing in a liquid/solid phase transition\cite{deVries:2014hq}.
 
Collectively, the in-plane magnetometry and transport measurements establish that we have successfully tuned exchange interactions in epitaxial Fe\textsubscript{0.52}Rh\textsubscript{0.48}/MgO(001) films such that, under ambient conditions, the layers exhibit bistable magnetic order. Switching between AF and FM states can be achieved by heating and cooling over a practical temperature range, accessible to Peltier devices. Next, we leverage the designed magnetic bistability of our films to demonstrate magnetic patterning through the local photothermal control of exchange interactions. 

\begin{figure*}[!bt]
\includegraphics[width=0.97\textwidth]{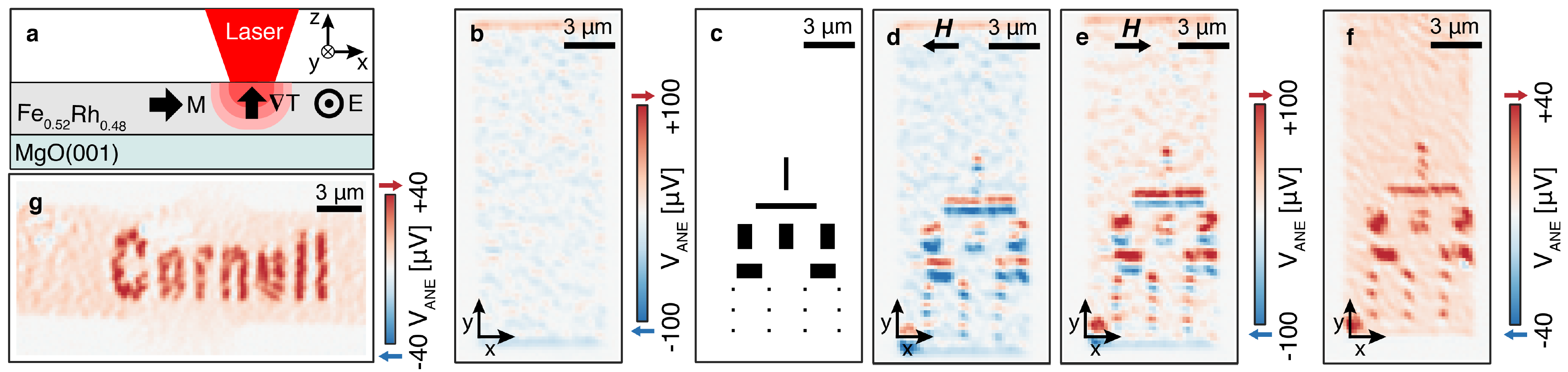}
\caption{\label{fig:ANE} 
\textbf{Local photothermal control of phase transition for on-demand rewritable magnetic patterning at room temperature.}
\textbf{a}, Schematic of ANE magnetic microscopy used to write and read magnetic patterns in photolithographically defined devices containing room-temperature magnetically tuneable epitaxial Fe\textsubscript{0.52}Rh\textsubscript{0.48} layers.
\textbf{b}, ANE image of a device initialised in the AF phase. Red (blue) regions represent moments pointing right (left). The contrast is due to uncompensated moments and residual ferromagnetism. 
\textbf{c}, The digital mask used to demonstrate photothermal magnetic patterning. 
Anomalous Nernst images collected at \textbf{d} +1 kOe and \textbf{e} -1 kOe fields (applied along $x$) from the same region in \textbf{b} after writing the pattern in \textbf{c}. Both magnetic contrast and anisotropic transport effects contribute to image formation. 
To isolate the magnetic signal, the half-difference of \textbf{d} and \textbf{e} is computed; the result is shown in \textbf{f}. FM regions (dark red) are clearly discernible from the AF background (light red). 
\textbf{g}, Another half-difference image of letters spelling out the authors' affiliation.
}
\end{figure*}

To interrogate the local magnetic order, we employ a microscope setup\cite{Bartell:2015hp} based on the anomalous Nernst effect (ANE), the thermal analogue of the anomalous Hall effect. A magnetic conductor with magnetisation $\vec{M}(\vec{r})$ subjected to a thermal gradient $\vec{\nabla} T(\vec{r})$ produces an electric field $\vec{E}_\textrm{ANE}(\vec{r})$ given by
\begin{equation}
        \vec{E}_\textrm{ANE}(\vec{r}) = -N \vec{\nabla} T(\vec{r}) \times \mu_o \vec{M}(\vec{r}),
\end{equation}
in which $N$ is the anomalous Nernst coefficient of the material and $\mu_o$ is the permeability of free space. Thermal gradients of approximately $0.15$ K/nm (maximum temperature difference $\sim$4 K) are produced primarily along the out-of-plane direction of our $\sim$35-nm-thick film by focusing a mode-locked Al\textsubscript{2}O\textsubscript{3}:Ti laser pulse ($\lambda = 785$ nm wavelength) with a fluence of 0.8 mJ/cm\textsuperscript{2} onto a diffraction-limited spot on the sample surface (see Fig. \ref{fig:ANE}\textbf{a}). The laser spot is then rastered across photolithographically defined 10 $\mu$m $\times$ 30 $\mu$m device structures; in this geometry, the resulting $\vec{E}_\textrm{ANE}$ engenders a voltage $V_\textrm{ANE}$ proportional to the in-plane component of $\vec{M}$ locally perpendicular to the device channel ($M_x$ in Figs. \ref{fig:ANE}\textbf{a}-\textbf{g}). 

Figure \ref{fig:ANE}\textbf{b} shows a representative $V_\textrm{ANE}$ map obtained at room temperature from a device initialised in the AF phase. The weak contrast observed is a combination of unpinned uncompensated moments that rotate with applied magnetic field \cite{Fan:2010gu} and pinned uncompensated moments that are strongly exchange-coupled to the bulk N\'eel order. The uncompensated moments are consistent with the 0.2 $\mu_B$/f.u. remanent magnetisation detected in the AF regime using magnetometry (Fig. \ref{fig:MH}\textbf{c}) and are characterised in detail in ref. \cite{Gray:ctOFlBwo}.

To switch the magnetic order, we increase the laser fluence ten-fold to 8 mJ/cm\textsuperscript{2}. Pulsed laser heating at this fluence causes a peak temperature increase in the Fe\textsubscript{0.52}Rh\textsubscript{0.48} of $\sim$40 K (estimated from finite-element simulations), which locally induces the FM phase while globally maintaining the sample at room temperature and in the AF state. Since our films are engineered to display magnetic bistability at room temperature, the induced FM regions persist even after photoheating when the region cools back to ambient temperature. We can therefore write FM regions at high fluence and image them without perturbing the written pattern at low fluence.

We demonstrate magnetic writing using the test pattern shown in Fig. \ref{fig:ANE}\textbf{c}, which includes rectangles of varying aspect ratio as well as pixel-sized dots for determining the minimum writing resolution. Figure \ref{fig:ANE}\textbf{d},\textbf{e} are ANE images collected from a patterned area with 1 kOe fields applied along the $-x$ and $+x$ directions, respectively. Written regions exhibit neighbouring positive (red) and negative (blue) contrast resembling dipoles. The contrast contains two components. The first component arises from spatial inhomogeneity in the thermal conductivity, which is higher in the FM phase of Fe\textsubscript{0.52}Rh\textsubscript{0.48} than in the AF phase. Near an in-plane AF/FM boundary, there is an imperfect cancellation of the in-plane charge Seebeck electric field along $+y$ and $-y$.  The sign of the resulting Seebeck voltage depends on which side of the thermal discontinuity the laser is focused, leading to a strong dipole-like feature.

The second component giving rise to the contrast in Fig. \ref{fig:ANE}\textbf{d},\textbf{e} is the ANE. To isolate this signal from the non-magnetic charge Seebeck response, we compute the half-difference of images Fig. \ref{fig:ANE}\textbf{d},\textbf{e}, which preserves features of the signal that switch in a magnetic field while eliminating features that are non-magnetic. The result, presented in Fig. \ref{fig:ANE}\textbf{f}, shows FM regions (dark red) within an AF background (light red). The light red contrast in the unwritten regions represents unpinned uncompensated moments. The observed FM shapes are consistent with the generating pattern and exhibit features with sub-micron dimensions, comparable to the laser spot size. To show that FM regions of arbitrary shape above this resolution can be patterned, we write and image another pattern with letters spelling out the authors' affiliation (Fig. \ref{fig:ANE}\textbf{f}). 

After FM writing, the film can be reset to the AF state by cooling with a Peltier device. In contrast to present magnetic writing procedures, which employ lithography, ion irradiation\cite{Kim:2012cm,Bali:2014dl} and implantation\cite{Fassbender:2008fe}, our process is fully repeatable and supports multiple write/erase cycles. Subsequent imaging reveals no permanent changes associated with the writing process. We envisage combining the processes discussed here with commercially mature heat-assisted magnetic recording technologies\cite{Stipe:2010ce} to achieve high-fidelity patterning with a resolution in the tens of nanometre range. This could promote novel competitive memories as well as facilitate prototyping of magnetic cellular automata\cite{Imre:2006is}, magnonic crystals\cite{Chumak:2015fa} and artificial spin-ice lattices\cite{Louis:2018bh}.

\section{Acknowledgements}

The authors thank K. Palmen and W. Zander for their help performing RBS measurements. A.B.M., Y.T. and D.G.S. acknowledge support in part by the Semiconductor Research Corporation (SRC) under nCORE tasks 2758.001 and 2758.003 and by the NSF under the E2CDA programme (ECCS-1740136). I.G. was supported by the Cornell Center for Materials Research with funding from the NSF MRSEC programme (DMR-1719875). J.B. acknowledges support by the AFOSR (FA9550‐14‐1‐0243). Materials synthesis was performed in a facility supported by the National Science Foundation (Platform for the Accelerated Realization, Analysis and Discovery of Interface Materials (PARADIM)) under Cooperative Agreement No. DMR-1539918. Substrate preparation was performed in part at the Cornell NanoScale Facility, a member of the National Nanotechnology Coordinated Infrastructure (NNCI), which is supported by the NSF (Grant No. ECCS-1542081).     

\section{Methods}

\subsection{Film growth}
Epitaxial Fe\textsubscript{1-$x$}Rh\textsubscript{$x$} films are grown using molecular-beam epitaxy to thicknesses of $\sim$35 nm on single-crystalline (001)-oriented MgO substrates in a Veeco GEN10 system with a base pressure of $1\times 10^{-8}$ Torr. Iron (99.995\% pure) and rhodium (99.95\% pure) species are simultaneously supplied to the growth surface from independent effusion cells. Molecular fluxes are calibrated using x-ray reflectivity (XRR) and quartz crystal microbalances and configured to produce films with rhodium fractions $x$ equal to 0.48. A substrate temperature of 420 $^\circ$C (estimated from a thermocouple in indirect contact with the growth surface and concealed from incident molecular fluxes) is employed for film growth and subsequent half-hour-long \emph{in situ} anneals. The anneal, which is performed immediately after film deposition, is designed to help order bcc Fe\textsubscript{1-$x$}Rh\textsubscript{$x$} alloys into the B2 CsCl-structure intermetallic with iron and rhodium residing on distinct positions of the two-atom basis.

\subsection{Compositional measurements}
Fe\textsubscript{1-$x$}Rh\textsubscript{$x$} film compositions are determined using RBS using a probe comprised of 1.4 MeV He$^+$ ions. The scattering geometry is defined by incident angle $\alpha = 7^\circ$, exit angle $\beta = 163^\circ$ and scattering angle $\theta = 170^\circ$. Spectra are integrated to a total accumulated ion dose of 15 $\mu$C. Film chemistry is determined by quantifying the area under iron and rhodium spectral features using an established procedure\cite{Petrov:1983fz}.

\subsection{Structural characterisation}
X-ray-based measurements are performed using a four-circle Philips X'pert MRD diffractometer operated with Cu$K_{\alpha1}$ radiation of wavelength $\lambda$ = 0.15406 nm ($\Delta\lambda/\lambda = 10^{-4}$). The incident beam optics consists of a four-bounce Ge 220  monochromator and a programable 0.125-mm-thick Ni attenuator. For XRR and XRD scans, a 1/16$^\circ$ divergence slit and a Xe proportional detector is employed as receiving optics. For $\omega$-rocking curve measurements, the divergence slit is replaced with a Ge 220  triple-axis analyzer crystal, proving an angular resolution of 12 arc-sec.

BF-TEM images and SAED patterns are collected in an FEI F20 transmission electron microscope with a field-emission source operated at 200 kV. The specimen foils are prepared near the MgO(010) zone axis by cutting vertical film sections in an FEI Strata 400 DualBeam. Initial milling is done using a 30 keV Ga\textsuperscript{+} focused ion beam. For final polishing, the ion energy is reduced to 5 keV.

\subsection{Magnetic characterisation}
The magnetic order of as-deposited Fe\textsubscript{1-$x$}Rh\textsubscript{$x$} layers is investigated in a Quantum Design physical property measurement system (PPMS). Temperature-dependent transport measurements are performed using the van der Pauw geometry with pressed-indium contacts by incrementally cycling the temperature in 5 K steps in the 200--395 K range. Magnetisation $\vec{M}$ vs. applied magnetic field $\vec{H}$ data is collected over the same temperature window by equipping the PPMS setup with a VSM module and orienting the sample such that Fe\textsubscript{1-$x$}Rh\textsubscript{$x$}$[110] \mid\mid \vec{H}$. At each temperature set point, the sample magnetisation is recorded while the magnetic field is swept between $\pm200$ Oe. 

\subsection{Magnetic imaging}
To write and image ferromagnetic patterns, we use an anomalous Nernst effect microscope.\cite{Bartell:2015hp} In this technique the local sample magnetization is transduced into an electrical voltage via a local thermal gradient. We generate local thermal gradients using a pulsed Coherent Mira 900 Al\textsubscript{2}O\textsubscript{3}:Ti laser tuned to 780 nm wavelength. We employ 3-ps-wide pulses and a repetition rate of 76 MHz (13 ns period). The laser is focused to a diffraction-limited 650 nm-diameter spot using a 0.90 numerical aperture microscope objective. We raster the laser using a 4f optical path in combination with a voice coil-controlled fast-steering mirror.  To detect $V_\textrm{ANE}$, we first collect the laser-induced voltage pulse train through a coplanar waveguide into a microwave transmission line and amplify the pulses by 40 dB with 0.1-3 GHz bandwidth. The pulse train is then sent to the radio-frequency port of a DC-12 GHz electrical mixer, where it is mixed with a 600 ps-wide pulse train from an arbitrary waveform generator that is referenced to the laser repetition rate. The mixer output voltage, $V_\textrm{ANE}$, is measured with a lock-in amplifier referenced to intensity modulation of the light with a photoelastic modulator. We estimated the temperature change induced by the laser using resistivity measurements in conjunction with time-domain finite element calculations performed in the COMSOL Multiphysics software package. 

\section{Author contributions}

A.B.M conceived and coordinated the study, synthesized the films and characterized their physical properties. I.G. fabricated the devices and performed the magnetic writing and imaging. Y.T. carried out the VSM analyses. J.S. conducted the RBS investigations and data analyses. D.W. performed the electron microscopy.  All authors contributed to the discussion and writing of the manuscript.

\bibliographystyle{naturemag}

\end{document}